\documentclass[conference]{IEEEtran}
\IEEEoverridecommandlockouts

\usepackage{cite}
\usepackage{amsmath,amssymb,amsfonts}
\usepackage{algorithmic}
\usepackage{graphicx}
\usepackage{textcomp}
\usepackage{xcolor}
\usepackage[hyphens]{url}
\usepackage{microtype}
\usepackage{booktabs}
\usepackage{circledsteps}
\usepackage{multirow} 
\usepackage{hyperref}
\usepackage{float}
\usepackage{graphicx}
\usepackage{subcaption}

\def\BibTeX{{\rm B\kern-.05em{\sc i\kern-.025em b}\kern-.08em
    T\kern-.1667em\lower.7ex\hbox{E}\kern-.125emX}}
\begin{document}

\pdfpagewidth=8.5in
\pdfpageheight=11in

\newcommand{\iscasubmissionnumber}{590}

\pagenumbering{arabic}

\title{Beyond Uniform Experts: Cost-Aware Expert Execution for Efficient Multi-Device MoE Inference}
\author{\textbf{Hui Zang, Pengfei Xia, Hong Liu, Jiajia Chu, Tuo Hao,  Minghao Chen, 
Rui Zhang, Ziyang Zhang}\\ Huawei Technologies Ltd}

\maketitle
\thispagestyle{plain}
\pagestyle{plain}

\begin{abstract}
Mixture-of-Experts (MoE) architectures enable language models to achieve unprecedented scale via sparse activation. However, their inference performance is often limited by data movement bottlenecks. Two coupled challenges exacerbate this limtation: (1) Importance-Agnostic Cost: Low-contribution experts incur nearly uniform memory and transfer costs, resulting in a low cost-to-benefit ratio and wasting critical bandwidth; (2) System-Level Imbalance: Multi-device deployments are universally bottlenecked by the slowest device, meaning that local reductions on one device may yield no improvement in end-to-end latency. We propose Cost-Aware Expert Execution (CAEE), a hardware-guided runtime framework that jointly optimizes for token-level expert importance and system-level execution cost. CAEE uses lightweight, calibrated cost models to estimate hardware overhead, selectively prunes low-importance, high-cost experts, and redistributes their contributions via a low-overhead compensation mechanism, avoiding extra data movement. Evaluations on the 671B DeepSeek-R1 model show that CAEE can reduce end-to-end inference latency by 8\%-18\% across diverse deployment settings, including expert offloading and on-device execution on multi-device systems, while maintaining a model accuracy drop of less than 1\%.
\end{abstract}

\section{Introduction}
Large Language Models (LLMs) \cite{brown2020language,openai2024gpt4,grattafiori2024llama3} have driven transformative progress in AI. However, their growing scale, now often exceeding hundreds of billions of parameters, introduces significant challenges for efficient deployment. Mixture-of-Experts (MoE) \cite{jiang2024mixtral,deepseekai2025deepseekv3} architectures alleviate the computational scaling problem via sparse activation, where only a small subset of $k$ experts processes each input token. While this achieves computational sparsity, it fails to address the memory and data movement burden: all $N$ expert parameters must remain accessible to support dynamic, token-level routing decisions. As a consequence, on modern accelerators, the data movement often emerges as the dominant bottleneck for MoE inference throughput across the memory hierarchy \cite{hwang2024pre,davies2025efficient}.

\begin{figure}[t]
\centering
\includegraphics[width=\linewidth]{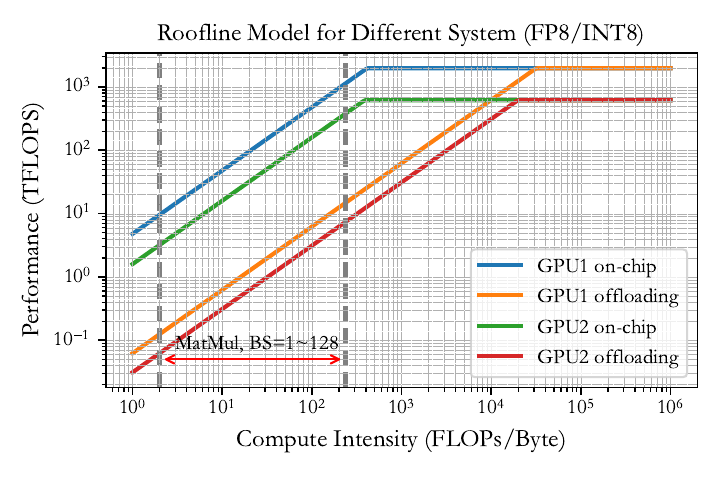}
\caption{Roofline analysis of MoE expert MatMul operations on two types of GPU. Within the range of batch sizes from 1 to 128, it is evident that the MoE inference is primarily dominated by data movement.}
\label{fig:roofline}
\end{figure}

\figurename~\ref{fig:roofline} illustrates the roofline analysis of MoE expert matrix multiplications (MatMuls) on two types of GPUs. Across a practical range of batch sizes from 1 to 128, the arithmetic intensity of the expert computation remains low, confirming that MoE inference is memory-bound rather than compute-bound. This bottleneck is severely amplified in scenarios involving expert offloading \cite{eliseev2023fast,tang2024hobbit}, where inactive experts must be swapped between host memory and device memory via interconnects like PCIe or CXL. With limited link bandwidths, typically in the tens of GB/s \cite{pcie,cxl}, the transfer latency restricts the overall throughput long before the device's computational resources are saturated, rendering the execution transfer-bound.

Mitigating this persistent data movement bottleneck requires a holistic approach that goes beyond simply reducing the number of activated experts. It mandates a joint consideration of the individual execution cost of each expert and its non-linear impact on the overall system performance. In practice, two coupled inefficiencies arise:
\begin{itemize}
\item Importance-Agnostic Cost (C1): Experts exhibit highly non-uniform token-level importance, yet they incur a nearly uniform runtime cost when activated. Every activated expert, regardless of whether it contributes $50\%$ or just $5\%$ to the final output logits, triggers a full memory access or a complete Host-to-Device (H2D) parameter transfer, wasting valuable memory bandwidth and interconnect resources.
\item System-Level Imbalance (C2): In multi-device systems, the end-to-end latency is dictated by the slowest device's completion time. The uneven, dynamic activation of experts across devices leads to load imbalance. Therefore, reducing the cost of a single expert on an underutilized device may yield little improvement in the end-to-end latency. Only by addressing the cost on the bottlenecked device or link can the system performance be improved.
\end{itemize}

These challenges motivate the development of a new runtime framework that precisely aligns the expert's importance with its actual hardware execution cost on both per-expert and multi-device dimensions.

Existing methods partially address these problems. Static pruning and compression\cite{chen2022task,he2024demystifying,xie2024moe} reduce the memory footprint and data movement by removing experts, but they ignore dynamic, token-level importance, which lowers model capacity. Dynamic routing approaches \cite{aghdam2024moe,li2023adaptive,zhong2024adapmoe} consider expert importance but typically ignore per-expert hardware cost or fail to evaluate system-wide, no-linear straggler impact. Moreover, most previous studies \cite{cao2025moe,eliseev2023fast,song2024promoe,kamahori2024fiddler} have typically been designed for single scenarios, leaving the effectiveness in both offloading and on-device configurations in multi-device systems untested.

Motivated by the challenges, we propose Cost-Aware Expert Execution (CAEE), a  hardware-guided runtime framework. CAEE refines expert execution by prioritizing cost reduction at the system's true bottleneck. Our main contributions are:
\begin{itemize}
\item Lightweight Multi-Device Cost Modeling: We develop a pragmatic cost model calibrated through brief offline profiling that estimates the per-expert data movement cost $\widetilde{C}(e,d)$, and crucially, a max-aggregation layer cost $F_{\text{cost}}$, which explicitly captures the system-level straggler effect across multiple devices $\mathcal{D}$.
\item Cost-Aware Pruning Strategy: We formulate the expert selection as an optimization problem, seeking to minimize the system-level cost $F_{\text{cost}}$ while bounding the removed importance $F_{\text{imp}}$. This strategy selectively bypasses only the high-cost, low-importance experts that contribute to the system's overall latency.
\item Low-overhead Compensation: To maintain accuracy, we introduce a mechanism that redistributes the contributions of pruned experts only to the remaining already-active or low-cost experts $\mathcal{A}_{\text{trans}}$. This ensures negligible accuracy degradation without incurring any additional memory access or data movement overhead.
\end{itemize}

These techniques allow CAEE to improve per-expert efficiency and reduce system-level stragglers, increasing end-to-end inference performance across both on-device and offloaded deployments. We evaluate CAEE on the 671B DeepSeek-R1 \cite{deepseekai2025deepseekr1} model across multiple settings and datasets. CAEE achieves a 8\%–18\% reduction in end-to-end inference latency with $< 1\%$ accuracy impact, demonstrating its effectiveness under diverse hardware bottlenecks.

\section{Preliminaries}
\subsection{MoE and Data Movement Bottleneck}
MoE architectures, exemplified by models like DeepSeek-R1\cite{deepseekai2025deepseekr1} and Qwen3\cite{yang2025qwen3}, extend the scalability of LLMs by introducing conditional computation. A router dynamically selects a small subset $\mathcal{K}_t$ of $k$ experts from $N$ total experts to process an input token $t$. The output $y_t$ is a weighted sum of the selected expert outputs: $$y_t = \sum_{e\in \mathcal{K}_t} G(t)_e \cdot E_e(t)$$ where $E_e(t)$ denotes the expert processing, $G(t)_e$ is the routing score for expert $e$ and $\sum G(t)_e = 1$. The Top-$k$ constraint ensures computational sparsity. 

However, the memory requirement scales linearly with $N$. Since $N$ can be in the hundreds for large MoE models, the memory footprint is immense. For example, DeepSeek-V3 \cite{deepseekai2025deepseekv3}, with 671 billion parameters, requires 1 TB of memory under FP16 quantization. The another core challenge arises during batched inference, where dynamic activation concurrently engages a substantial number of experts. This mandates the data movement of a large fraction of expert weights, either through high-frequency on-device memory accesses or via offloading from a slower host memory tier, e.g., DRAM, NVMe. As a result, parameter fetching latency rapidly becomes the primary bottleneck, overshadowing the computational gains of sparsity.

\subsection{Multi-Device Systems}
\begin{figure}[t]
\centering
\includegraphics[width=\linewidth]{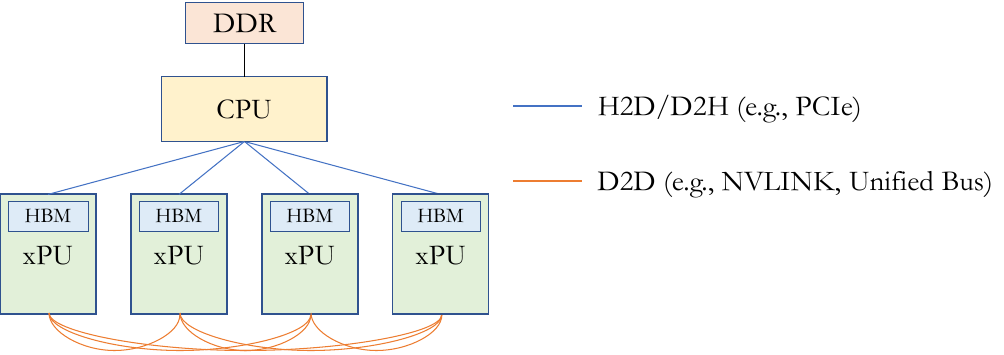}
\caption{A logical topology diagram for a single-node multi-device server, where xPU represents computation devices such as GPU.}
\label{fig:multi-device}
\end{figure}

In modern computing environments, multi-device servers are standard for LLM inference and training. As shown in \figurename~\ref{fig:multi-device}, these systems typically feature host servers equipped with multiple computation devices, each possessing its own HBM and computational units. The key components include:
\begin{itemize}
\item Device Interconnects: High-speed, low-latency links, e.g., NVLINK, connecting the devices for fast data exchange in parallel execution.
\item Host-Device Interconnects: Slower links, e.g., PCIe \cite{pcie}, CXL \cite{cxl}, connecting the devices to the host CPU and DRAM for system-wide memory access.
\end{itemize}

This architecture enables Expert Parallelism (EP) \cite{rajbhandari2022deepspeed} for MoE, where the $N$ experts are partitioned across the $\mathcal{D}$ devices. A token's execution flow involves: first, token dispatch, as the router decides which device $d$ is responsible for computing the token based on the required expert $e$; second, expert activation, ensuring device $d$ has expert $e$'s parameters in its HBM; third, computation, where expert $e$ processes the token; and finally, result aggregation, gathering results across devices. The central challenge is that dynamic and unpredictable routing leads to uneven memory access and data transfer patterns across the devices and links \cite{huang2023towards,deepseekai2025deepseekv3}.

\begin{figure*}[t]
\centering
\includegraphics[width=1.0\linewidth]{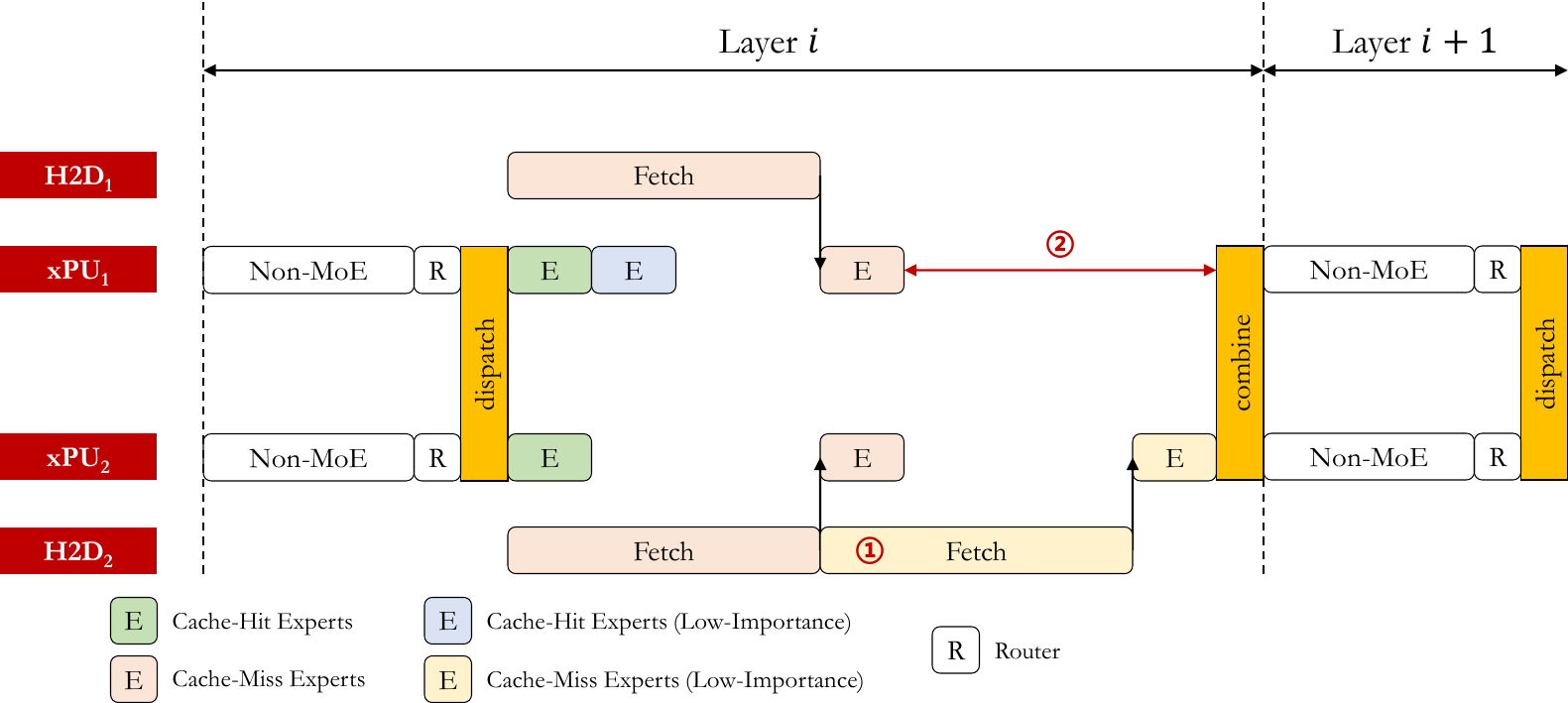}
\caption{Illustration of the two key challenges in offloading MoE models on multi-device systems. (\Circled{1}) For cache-miss experts, the retrieval cost is high regardless of the token's importance score; redirecting tokens away from low-importance, cache-miss experts can save H2D bandwidth. (\Circled{2}) Uneven expert activation leads to idle devices and underutilized H2D bandwidth, reducing expert transmissions on idle links may have little to no effect on improving end-to-end performance.}
\label{fig:challenges}
\end{figure*}

\section{Analysis and Challenges}

In this section, we analyze two fundamental challenges, that arise when deploying MoE models on multi-device systems. For clarity, we use the expert offloading scenario \cite{eliseev2023fast,tang2024hobbit} as a running example, because it amplifies data-movement behavior and makes the issues visible in practice. However, these challenges are not unique to offloading: they stem from the general data-movement–bound nature of MoE inference and persist across on-device deployments. Offloading is therefore an illustrative case rather than the only scenario of concern. \figurename~\ref{fig:challenges} highlights these two challenges.

\subsection{C1: Importance-Agnostic Expert Transfer}
As depicted in \figurename~\ref{fig:challenges} (Yellow Box \Circled{1}), the execution cost of activating an expert is largely independent of its routing score $G(t)_e$. This behavior exposes a mismatch:
\begin{itemize}
\item Uniform Transfer Cost: When an expert's parameters are needed for a token and are not currently in the device's HBM (a cache-miss), the system must execute a full transfer of its entire weight tensor $S_e$ from the slower memory tier. This transfer operation consumes a fixed and substantial amount of memory or link bandwidth $B_d$.
\item Highly Variable Importance: For a Top-$k$ routing manner, the routing scores $G(t)_e$ can differ by orders of magnitude. The $k$-th selected expert might have a score of 5\%, while the top expert could score over 50\%. Both, if cache-misses, incur the same transfer cost.
\end{itemize}

The result is a fundamental mismatch: high-cost, low-importance experts consume scarce data movement resources, creating a low cost-to-benefit ratio. For example, in an offloading system, directing a token away from a low-importance, cache-miss expert can save tens of milliseconds of H2D transfer latency, which is far more impactful than rerouting tokens away from a high-importance, cache-hit expert. 

\subsubsection{Quantitative Analysis}
\begin{figure}[t]
\centering
\includegraphics[width=\linewidth]{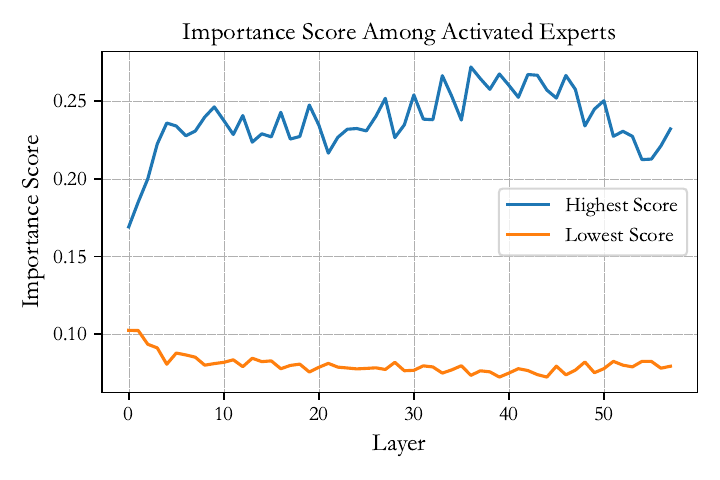}
\caption{Variation in expert importance across layers. For each layer, we compute the average of the highest and lowest importance scores among the experts activated during 16 batch sizes and 100 decoding processes. The differences in expert importance were smaller in earlier layers, but after Layer 10, the scores of the highest-importance experts were typically 3 to 4 times higher than those of the lowest-importance experts. Despite this divergence, activating either expert incurs the same transfer cost, illustrating the mismatch that underlies C1.}
\label{fig:c1_example}
\end{figure}

We analyze runtime statistics from the 671B DeepSeek-R1 model on an 8 device system with expert offloading. \figurename~\ref{fig:c1_example} plots the layer-wise average highest and lowest importance scores among activated experts. The divergence is significant: after Layer 10, the maximum score is typically 3$\times$ to 4$\times$ greater than the minimum score. This contrast confirms the high variability in expert importance. Since the data movement cost for both experts is the same, this divergence is the empirical evidence for the C1 mismatch.

\subsection{C2: Data Movement Imbalance Across Devices}
As highlighted in \figurename~\ref{fig:challenges} (Red Line \Circled{2}), in multi-device EP deployments, the distribution of expert activation and data movement is rarely uniform. We can observe that:
\begin{itemize}
\item Uneven Workload: Dynamic routing based on token semantics leads to an unpredictable distribution of expert demands. One device might handle tokens that primarily hit cached experts, while another device is tasked with tokens that trigger multiple cache-misses, leading to its H2D link being fully saturated.
\item Straggler Effect: The system performance is capped by the slowest device, i.e., $\text{Latency}_{\text{total}} = \max_{d \in \mathcal{D}} (\text{Latency}_d)$. Therefore, reducing $\text{Latency}_d$ on an already fast device does not actually improve $\text{Latency}_{\text{total}}$. Only reducing the cost on the busiest device or link can yield system-wide performance improvement.
\end{itemize}

This confirms that naive, uniform pruning based on global metrics is ineffective and potentially harmful. A cost-aware pruning strategy must explicitly minimize the maximum cost across all devices.

\subsubsection{Quantitative Analysis}
\begin{figure}[t]
\includegraphics[width=\linewidth]{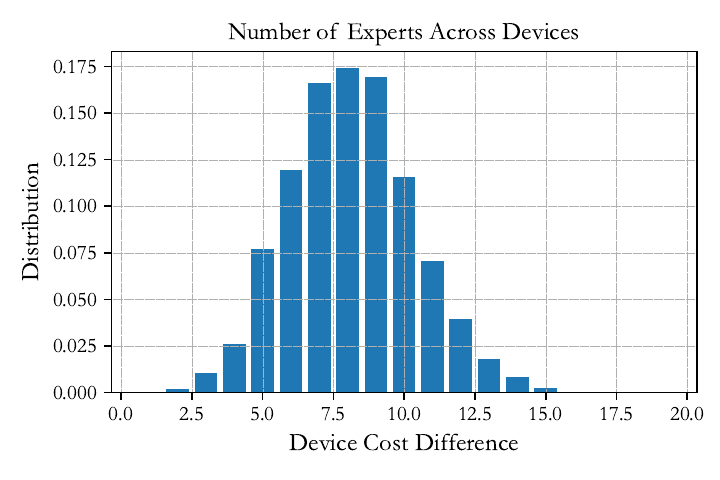}
\caption{Distribution of device-level expert-transfer imbalance. For each inference run, we calculate the number of experts migrated across all 8 devices and record the difference between the busiest and the least busy devices. A difference of 0 indicates perfect balance. Most runs exhibit significant imbalance, revealing the inherent bias driving C2.}
\label{fig:c2_example}
\end{figure}

\begin{figure}[t]
\includegraphics[width=\linewidth]{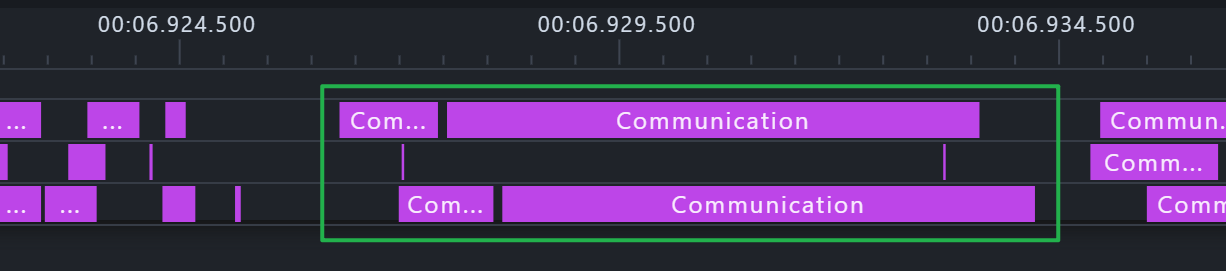}
\caption{Communication imbalance observed in profiling data. Among the three devices, the longest communication time determines the overall latency.}
\label{fig:imbalance}
\end{figure}

\figurename~\ref{fig:c2_example} aggregates the number of expert transfers across the 8 devices. The results show that: perfect balance, gap = 0, is rare, while significant imbalance, gap of 6 to 10 transfers, occurs frequently. This means that, for a given inference step, one device or link is handling up to 10 more expert transfers than another. Another practical profiling data example is shown in \figurename~\ref{fig:imbalance}.

This imbalance is the empirical evidence for C2, proving the existence of the straggler problem. This underscores the necessity of a cost model that uses max-aggregation to target the true performance bottleneck.

\section{System Design}
\begin{figure*}[t]
\centering
\includegraphics[width=1.0\linewidth]{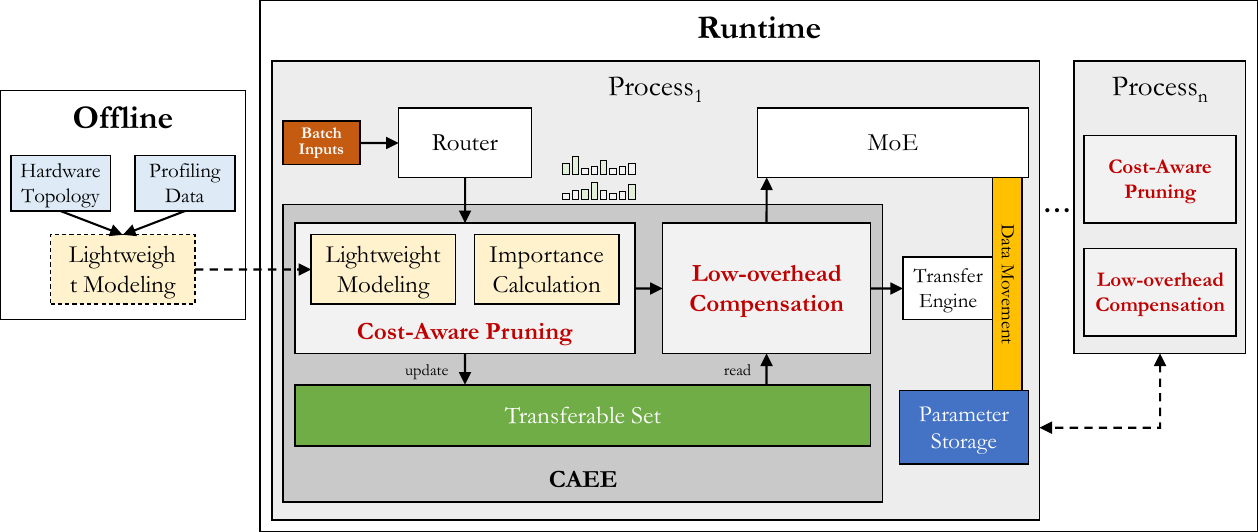}
\caption{Overview of the CAEE framework. CAEE operates within the runtime of each device $\text{Process}_i$ in a multi-device system. The offline phase calibrates the lightweight modeling component using hardware topology and profiling data to estimate per-expert and per-system cost. At runtime, the cost-aware pruning module integrates the cost estimate and importance calculation to derive the final set of active experts, known as the transferable set $\mathcal{A}_{\text{trans}}$. The low-overhead compensation mechanism only reroutes the pruned contribution values to the expert nodes in the transferable set, thereby ensuring that the redistribution is completed without making any additional data movement requests to the transfer engine.}
\label{fig:system}
\end{figure*}

In this section, we detail the design of Cost-Aware Expert Execution (CAEE), which is executed as a runtime component on each device $\text{Process}_i$ to dynamically refine expert selection. The framework's goal is to transition from importance-only routing to a cost-aware routing decision that minimizes the system-level straggler latency. 

As illustrated in \figurename~\ref{fig:system}, CAEE operates in two phases: an offline phase for cost model calibration, and a runtime phase incorporating the core mechanisms: (1) hardware cost modeling, (2) cost-aware pruning, and (3) low-overhead compensation.

\subsection{Hardware Cost Modeling}
CAEE's hardware cost model is centered around the idea that the total layer latency $F_{\text{cost}}$ is determined by the maximum data movement cost across all devices.

\subsubsection{Per-Expert Data-Movement Cost}
For an expert $e$ assigned to device $d$, the cost $C(e,d)$ is based on the volume of data movement required. The core formulation unifies the cost regardless of the source memory: $$C(e,d) \propto \frac{S_e}{B_d} + \text{Overhead}_d(e)$$ where $S_e$ is the fixed parameter size of expert $e$, $B_d$ is the effective data-movement bandwidth associated with device $d$, e.g., HBM bandwitdh or PCIe link bandwidth, and $\text{Overhead}_d(e)$ is a constant term representing the scheduler overhead. 

\subsubsection{Layer-Level Cost via Max Aggregation}
To capture the C2 straggler effect, the layer cost $F_{\text{cost}}$ is defined by the maximum total cost incurred by any device in the system: $$F_{\text{cost}}(\mathcal{X}) = \max_{d \in \mathcal{D}} \left( \sum_{e \in \mathcal{E}_d} x_e \cdot R(e,d) \cdot C(e,d) \right)$$ Here, $\mathcal{X} = \{x_e\}$ is the set of expert activation decisions and $x_e=1$ if active, $\mathcal{E}_d$ is the set of experts assigned to device $d$, and $C(e,d)$ is the intrinsic parameter transfer cost. Crucially, $R(e,d)$ represents the dynamic memory residency status: for on-device inference, $R(e,d)$ is typically set to $1$ to reflect uniform memory access cost or HBM pressure; conversely, for expert offloading systems, $R(e,d) = 1$ if expert $e$ triggers an expensive H2D transfer, and $R(e,d) = 0$ if it is a cache-hit. By minimizing $F_{\text{cost}}(\mathcal{X})$, we directly target the throughput bottleneck by reducing the workload on the busiest device.

\begin{table*}[t]
\centering
\caption{Experimental setup for offloading and on-device deployments.}
\label{tab:exam_setup}
\begin{tabular}{c|ccccccccccc}
\toprule
Task & Device Num & LLM Engine & Deploy Strategy & R1 Thinking \\
\midrule
Offloading  & 8 & vLLM & MLA: DP1-TP8, MoE: EP8 & Disabled \\
On-device & 16 & MindIE & MLA: DP2-TP8, MoE: EP16 & Enabled \\
\bottomrule
\end{tabular}
\end{table*}

\subsubsection{Offline Profiling and Calibration}
To account for real-world non-idealities such as contention, varying clock speeds, and firmware overheads, CAEE performs a brief offline profiling phase. This calibrates the effective bandwidth $B_d$ and fixed overheads: $$\widetilde{C}(e,d) = \alpha_d \cdot \frac{S_e}{B_d} + \beta_d$$ where $\alpha_d$ is the bandwidth degradation factor and $\beta_d$ is the fixed transfer overhead. This calibration ensures the cost model's accuracy, making the online cost estimation $\widetilde{C}(e,d)$ highly lightweight, enabling real-time inference through simple table lookups and aggregation operations.

\subsection{Cost-Aware Pruning}
Our goal is to select a subset of experts from the originally activated ones, that is, $\mathcal{A} \subseteq \mathcal{A}_{\text{act}}$ that significantly reduces the system cost $F_{\text{cost}}$ while minimally impacting the total expert importance $F_{\text{imp}}$. We use $I(e)$ to denote the importance of expert $e$. The objective can be formulated as $$\text{minimize: } F_{\text{cost}}(\mathcal{X}) = \max_{d \in \mathcal{D}} \left( \sum_{e \in \mathcal{E}_d} x_e \cdot R(e,d) \cdot \widetilde{C}(e,d) \right)$$$$\text{subject to: } F_{\text{imp}}(\mathcal{X}) = \sum_{e=1}^{N} (1 - x_e) \cdot I(e) \le \varepsilon$$
where $\varepsilon$ bounds the acceptable accuracy degradation.

Due to the strict time requirements for inference speed, CAEE solves the constrained optimization problem using a efficient cost-invariant greedy heuristic that minimizes $F_{\text{cost}}$ while strictly bounding the importance loss. The process dynamically determines the final set of transferable experts $\mathcal{A}_{\text{trans}} \in \mathcal{A}_{\text{act}}$  via four steps:
\begin{enumerate}
\item Initial Expert Set Separation: We first partition the entire set of activated experts $\mathcal{A}_{\text{act}}$ into two groups based on a predetermined high importance threshold $\theta$: (1) Mandatory Set $\mathcal{A}_{\text{must}}$: Experts whose importance score $I(e)$ is above $\theta$ are critical for accuracy, i.e., $\mathcal{A}_{\text{must}} = \{ e \mid I(e) \ge \theta \text{ and } e \in \mathcal{A}_{\text{act}} \}$. (2) Candidate Set $\mathcal{A}_{\text{cand}}$: Experts whose scores are below $\theta$, representing potential candidates for removal, that is $\mathcal{A}_{\text{cand}} = E \setminus \mathcal{A}_{\text{must}}$.
\item Baseline Cost Determination: The initial baseline execution cost $F_{\text{base}}$ is calculated based solely on the mandatory experts $\mathcal{A}_{\text{must}}$, as they must be transferred or accessed:$$F_{\text{base}} = \max_{d \in \mathcal{D}} \left( \sum_{e \in \mathcal{A}_{\text{must}} \cap \mathcal{E}_d} R(e,d) \cdot \widetilde{C}(e,d) \right)$$
\item Cost-Invariant Expansion: To maximize accuracy without increasing the critical path latency, we incorporate experts from $\mathcal{A}_{\text{cand}}$. An expert $e \in \mathcal{A}_{\text{cand}}$ is promoted to $\mathcal{A}_{\text{must}}$ if and only if its inclusion does not increase the layer's total execution time above the baseline cost $F_{\text{base}}$, i.e., $$\text{if } \max_{d \in \mathcal{D}} \left( \sum_{e' \in \mathcal{A}_{\text{must}} \cup \{e\}} R(e,d) \cdot \widetilde{C}(e',d) \right) = F_{\text{base}} $$ $$ \text{ then } \mathcal{A}_{\text{must}}  \leftarrow \mathcal{A}_{\text{must}} \cup \{e\}$$ This key step effectively performs cost-aware expert expansion: it prioritizes retaining low-importance experts that happen to be cache-hits or reside on underutilized links, ensuring we maximize utility without contributing to the straggler effect.
\item Final Pruning and Expert Set Definition: After the expansion phase, the experts remaining in $\mathcal{A}_{\text{cand}}$ are officially designated as the pruned set, as their inclusion would increase the latency $F_{\text{cost}}$ beyond $F_{\text{base}}$. The final set of active, transferable experts is defined as $\mathcal{A}_{\text{trans}} = \mathcal{A}_{\text{must}}$ and $\mathcal{A}_{\text{trans}} \subseteq \mathcal{A}_{\text{act}}$. 
\end{enumerate}
This strategy is highly effective because it ensures the pruning decision is guided by the hardware cost at the bottleneck, i.e., maintaining $F_{\text{cost}} = F_{\text{base}}$, rather than relying on simple global importance scores. The resulting set $\mathcal{A}_{\text{trans}}$ is optimized for both speed and model accuracy.

\subsection{Low-overhead Compensation}
The low-overhead compensation mechanism is deployed to finalize the expert selection after the cost-aware pruning determines the set of active, transferable experts, $\mathcal{A}_{\text{trans}}$. This mechanism is designed for zero additional memory traffic, operating exclusively via fast, local data manipulation to maintain accuracy while respecting the pruning decision. The core principle is to constrain the final routing decision only to the experts remaining in $\mathcal{A}_{\text{trans}}$, ensuring that no memory-intensive data movement is initiated for any pruned expert.

For a given token $t$, the compensation process is simplified to the following two steps:
\begin{enumerate}
\item Masking of Pruned Experts: The scores of all experts $e$ outside the transferable set $\mathcal{A}_{\text{trans}}$ are effectively reset to zero, creating a masked score set $G_{\text{masked}}(t)$:$$G_{\text{masked}}(t)_e = \begin{cases} G(t)_e & \text{if } e \in \mathcal{A}_{\text{trans}} \\ 0 & \text{if } e \notin \mathcal{A}_{\text{trans}} \end{cases}$$ This masking mechanism ensures that the pruned expert models are completely disregarded during the final output aggregation process, while retaining the scores of all transferable experts.
\item Final Top-$k$ Selection: The final set of $k$ active experts $\mathcal{K}_{\text{final}}$ is selected by applying the standard Top-$k$ operation on the masked score set $G_{\text{masked}}(t)$:$$\mathcal{K}_{\text{final}} = \text{Top-}k(G_{\text{masked}}(t))$$This step effectively re-routes the token's contribution, which would have gone to the pruned experts, to the remaining highest-scoring experts within the low-overhead set $\mathcal{A}_{\text{trans}}$. This ensures that the model output remains dense and maximizes the utility provided by the cost-optimized set.
\end{enumerate}

\section{Evaluation}
We evaluate the proposed framework with two MoE deployments: the \textit{offloading} scenario and the \textit{on-device} scenario. 

\subsection{Experimental Setup}
\subsubsection{Models and Benchmarks}
We employ DeepSeek-R1 671B \cite{deepseekai2025deepseekr1} with W8 quantization as our representative MoE model. The model contains 671B total parameters with 256 experts, but activates only 37B parameters and 8 experts per token during inference. A key innovation is its Multi-Head Latent Attention (MLA) \cite{deepseekai2024deepseekv2}, which replaces standard attention to achieve a massive reduction in the Key-Value (KV) cache size.

To evaluate the impact of CAEE on model accuracy, we employ multiple open-source benchmarks across three domains: Knowledge (MMLU-shot5 \cite{hendrycks2020measuring}, CEval-shot5 \cite{huang2023c}), Math (GSM8K-shot5 \cite{cobbe2021training}), and Code (HumanEval \cite{chen2021evaluating}). For inference performance, we measure Time to First Token (TTFT) and Time Per Output Token (TPOT) under fixed concurrency, while adjusting the per-device concurrency according to each deployment scenario.

\subsubsection{Implementation Details}
To guarantee a diverse evaluation, the experiments are conducted across separate offloading and on-device deployments of DeepSeek-R1, each utilizing different hardware, inference frameworks, and routing strategies. \tablename~\ref{tab:exam_setup} outlines the corresponding experimental setup.

\begin{table*}[t]
\centering
\caption{Performance of CAEE in the expert offloading scenario.}
\label{tab:performance_offloading}
\begin{tabular}{c|cccccccc}
\toprule
 & \multicolumn{2}{c}{Baseline} & \multicolumn{2}{c}{$\theta_{\text{imp}}=0.9$} & \multicolumn{2}{c}{$\theta_{\text{imp}}=0.8$}  & \multicolumn{2}{c}{$\theta_{\text{imp}}=0.7$} \\
\cmidrule(lr){2-3}\cmidrule(lr){4-5}\cmidrule(lr){6-7}\cmidrule(lr){8-9}
& TTFT (s) & TOPT (s) & TTFT (s) & TOPT (s) & TTFT (s) & TOPT (s) & TTFT (s) & TOPT (s) \\
\midrule
Batch Size = 1 & 6.171 & 0.819 & 5.973 & 0.768 & 5.887 & 0.704 & 5.822  & 0.645 \\ 
Gain vs. Baseline  & /  & / & 3.31\%↓  & 6.23\%↓ & 4.60\%↓ & 14.04\%↓ & 5.66\%↓  & 21.25\%↓ \\ \midrule
Batch Size = 4  & 16.638 & 0.957 & 16.236 & 0.914 & 15.971 & 0.824 & 15.635 & 0.765 \\
Gain vs. Baseline  & / & / & 2.48\%↓ & 4.49\%↓ & 4.01\%↓ & 13.90\%↓ & 6.03\%↓ & 20.06\%↓ \\ \midrule
Batch Size = 8  & 27.961 & 1.024 & 27.381 & 0.979 & 26.863 & 0.849 & 26.477 & 0.796 \\ 
Gain vs. Baseline  & / & / & 2.12\%↓ & 4.39\%↓ & 3.93\%↓ & 17.09\%↓ & 5.31\%↓ & 22.27\%↓ \\ \midrule
Batch Size = 16 & 46.273 & 1.377 & 45.278 & 1.317 & 44.703 & 1.143 & 44.100 & 1.057 \\
Gain vs. Baseline & / & / & 2.20\%↓ & 4.36\%↓ & 3.39\%↓ & 16.99\%↓ & 4.70\%↓ & 23.24\%↓ \\
\bottomrule
\end{tabular}
\end{table*}

\begin{table*}[t]
\centering
\caption{Accuracy comparison of CAEE in the expert offloading scenario.}
\label{tab:accuracy_offloading}
\begin{tabular}{lcccc}
\toprule
Dataset & Baseline & $\theta_{\text{imp}}=0.9$ & $\theta_{\text{imp}}=0.8$ & $\theta_{\text{imp}}=0.7$ \\
\midrule
CEval & 0.8841 & 0.8796 (0.51\%↓) & 0.8789 (0.59\%↓) & 0.8662 (2.02\%↓) \\
MMLU  & 0.8513 & 0.8464 (0.58\%↓) & 0.8440 (0.86\%↓) & 0.8398 (1.35\%↓) \\
GSM8K & 0.9561 & 0.9529 (0.33\%↓) & 0.9498 (0.66\%↓) & 0.9404 (1.64\%↓) \\
\bottomrule
\end{tabular}
\end{table*}

\subsection{Expert Offloading}
\begin{figure}[t]
\includegraphics[width=\linewidth]{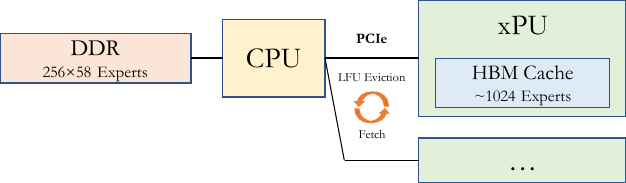}
\caption{Expert offloading system used in the experment.}
\label{fig:offloading}
\end{figure}

The expert offloading system evaluated in this work is depicted in \figurename~\ref{fig:offloading}. Experiments are conducted on a computing server featuring 8 xPUs and 1.5 TB of host memory. Since the DeepSeek-R1 model's 58 MoE layers (totaling 14848 experts) exceed the available device memory, each HBM caches only a subset of 1024 experts, managed by a Least Frequently Used (LFU) eviction policy. Each expert has a size of 43 MB, corresponding to weight dimensions of $3\times7168\times2048$.

The expert offloading scenario, where the H2D transfer dominates the critical path, affords greater flexibility than highly latency-sensitive on-device inference. This allows us to define the mandatory expert set, $\mathcal{A}_{\text{must}}$, using a more selective, importance-based retention strategy. For an input token, the Top-$K$ experts are sorted by their gating scores $G(t)_e$. We then dynamically populate $\mathcal{A}_{\text{must}}$ by retaining the minimal number of top-ranked experts ($m^*$) whose Cumulative Importance Ratio ($\text{CIR}$), defined as $\text{CIR}(m) = \sum_{i=1}^{m} G(t)_{e_i} / \sum_{j=1}^{K} G(t)_{e_j}$, satisfies a preset retention threshold $\theta_{\text{imp}}$, i.e., $\text{CIR}(m^*) \ge \theta_{\text{imp}}$. Experts not retained in $\mathcal{A}_{\text{must}}$ are subsequently subjected to the cost-aware pruning mechanism, which targets low-importance contributions that would otherwise incur high H2D transfer costs. We evaluate the utility-latency trade-off by setting $\theta_{\text{imp}} \in \{0.9, 0.8, 0.7\}$ in our experiments.

\subsubsection{Impact of Preset Retention Threshold $\theta_{\text{imp}}$}
Tables \ref{tab:performance_offloading} and \ref{tab:accuracy_offloading} present the performance and accuracy of CAEE. We observe the most pronounced gains in the TOPT metric, which governs steady-state throughput. By pruning of low-importance, high-cost experts, CAEE significantly reduces unnecessary per-token data transfers, successfully addressing the PCIe link saturation. TOPT reduction ranges from $4.36\%$ ($\theta_{\text{imp}}=0.9$) up to a maximum of $23.24\%$ ($\theta_{\text{imp}}=0.7$) at Batch Size 16. While TTFT also improves (up to $6.03\%$), the dominant benefit is in the recurrent TOPT phase.


\begin{figure}[ht]
\centering
\includegraphics[width=\linewidth]{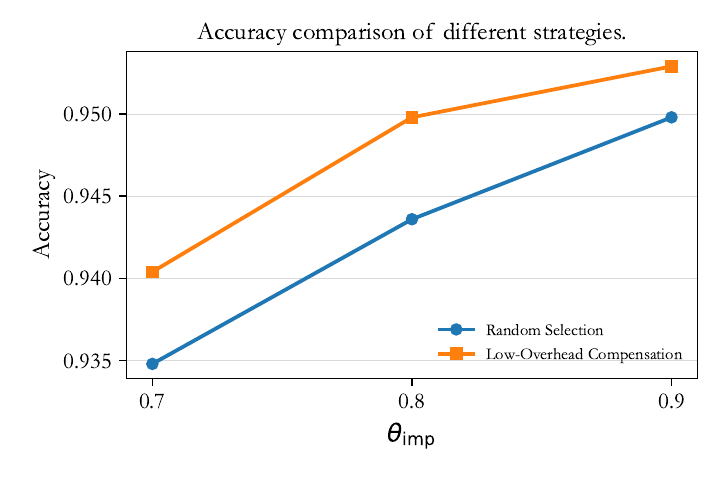}
\caption{Accuracy comparison between random selection and low-overhead compensation across different $\theta_{\mathrm{imp}}$ values.}
\label{fig:compensation}
\end{figure}

Crucially, these substantial performance gains are achieved with minimal impact on accuracy, validating our cost-aware pruning and low-overhead compensation. The configuration $\theta_{\text{imp}}=0.8$ presents the most favorable trade-off: it consistently yields TOPT reductions between $13.90\%$ and $17.09\%$ across all batch sizes, while restricting accuracy degradation to less than $1\%$ on MMLU ($0.86\%$) and GSM8K ($0.66\%$). The ability of CAEE to substantially improve throughput while maintaining high utility confirms its viability as a runtime optimization framework for transfer-bound MoE deployments.

\subsubsection{Ablation Study on Low-overhead Compensation}
When rerouting low-importance experts to those with smaller loading costs, we contrast a random selection strategy with the proposed low-overhead compensation mechanism; \figurename~\ref{fig:compensation} confirms that our method consistently yields superior accuracy on the GSM8K dataset, validating its effectiveness.

\subsubsection{Hardware Cost Observation}

\begin{figure*}[htbp]
\centering
\subfloat[Layer 2]{\label{fig:subfig1}\includegraphics[width=0.24\textwidth]{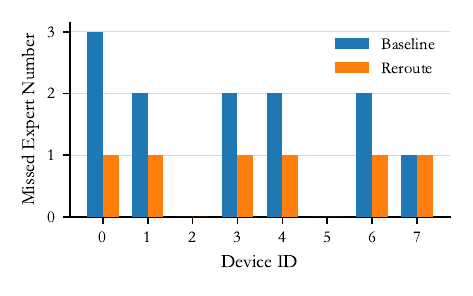}}
\subfloat[Layer 8]{\label{fig:subfig2}\includegraphics[width=0.24\textwidth]{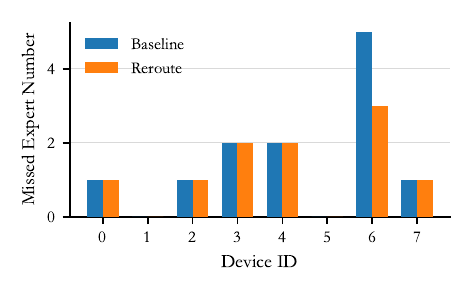}}
\subfloat[Layer 16]{\label{fig:subfig2}\includegraphics[width=0.24\textwidth]{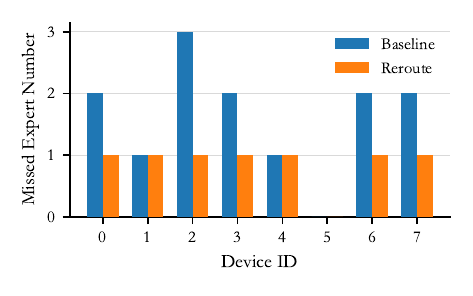}}
\subfloat[Layer 24]{\label{fig:subfig2}\includegraphics[width=0.24\textwidth]{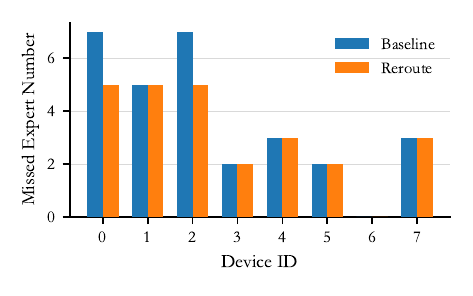}} \\
\subfloat[Layer 30]{\label{fig:subfig2}\includegraphics[width=0.24\textwidth]{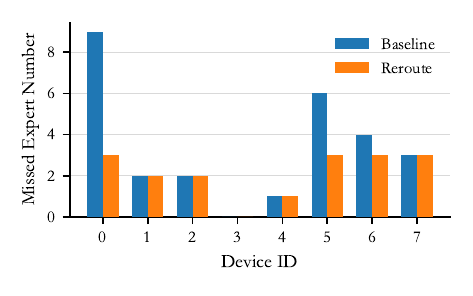}}
\subfloat[Layer 32]{\label{fig:subfig2}\includegraphics[width=0.24\textwidth]{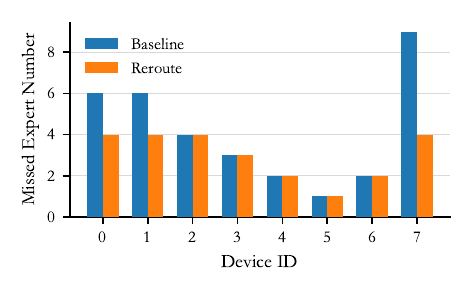}}
\subfloat[Layer 39]{\label{fig:subfig2}\includegraphics[width=0.24\textwidth]{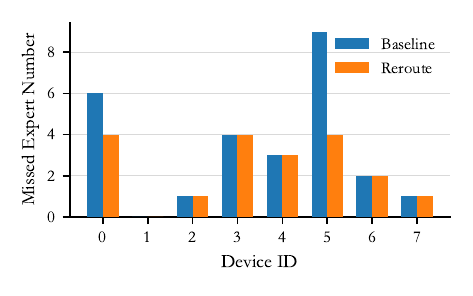}}
\subfloat[Layer 57]{\label{fig:subfig2}\includegraphics[width=0.24\textwidth]{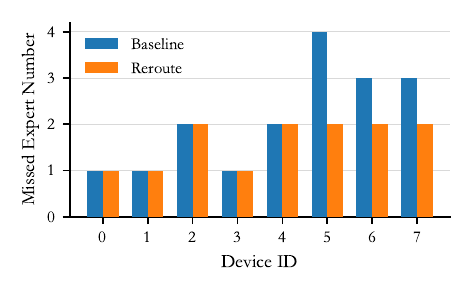}}
\caption{Distribution of expert transfer load across 8 devices on selected MoE layers. The results show that CAEE successfully mitigates the system-level straggler effect by actively reducing the transfer count on the busiest device, compared to the baseline.}
\label{fig:cost_observation}
\end{figure*}

\begin{table*}[t]
\centering
\caption{Performance of CAEE in the on-device inference scenario. TPOT (ms) is tested as a performance metric.}
\label{tab:on_device_performance}
\begin{tabular}{c|ccccccccccc}
\toprule
$\theta$ & Baseline & 0.30 & 0.40 & 0.50 & 0.60 
& 0.70 & 0.80 & 0.90 & 1.0 & 1.1 & 1.2 \\
\midrule
Batch Size = 64  & 59.29 & 58.46 & 56.08 & 55.72 & 54.58 & 53.15 & 52.30 & 51.66 & / & / & / \\
Gain vs. Baseline & / & 1.40\%↓ & 5.41\%↓ & 6.01\%↓ & 7.94\%↓ & 10.36\%↓ & 11.78\%↓ & 12.87\%↓ & / & / & / \\
\midrule
Batch Size = 128 & 68.78 & 68.92 & 66.63 & 65.70 & 64.55 & 62.95 & 62.18 & 61.66 & 60.79 & 60.33 & 59.42 \\
Gain vs. Baseline & / & 0.2\%↑ & 3.13\%↓ & 4.49\%↓ & 6.15\%↓ & 8.47\%↓ & 9.59\%↓ & 10.36\%↓ & 11.63\%↓ & 12.30\%↓ & 13.62\%↓ \\
\bottomrule
\end{tabular}
\end{table*}

\begin{table*}[t]
\centering
\caption{Accuracy comparison of CAEE in the on-device inference scenario.}
\label{tab:on_device_accuracy}
\begin{tabular}{cc|ccccccccccc}
\toprule
Dataset & BS 
& Baseline & 0.30 & 0.40 & 0.50 & 0.60 & 0.70 & 0.80 & 0.90 & 1.0 & 1.1 & 1.2 \\
\midrule
\multirow{2}{*}{CEval}
& 64  & 0.8952 & 0.8952 & 0.8952 & 0.8952 & 0.8952 & 0.8952 & 0.8952 & 0.8952 & / & / & / \\
& 128 & 0.8952 & 0.8952 & 0.8952 & 0.8952 & 0.8952 & 0.8952 & 0.8952 & 0.8952 & 0.8952 & 0.8952 & 0.8952 \\
\midrule
\multirow{2}{*}{MMLU}
& 64  & 0.8683 & 0.8683 & 0.8683 & 0.8683 & 0.8683 & 0.8683 & 0.8683 & 0.8683 & / & / & / \\
& 128 & 0.8683 & 0.8694 & 0.8694 & 0.8694 & 0.8694 & 0.8694 & 0.8694 & 0.8694 & 0.8694 & 0.8694 & 0.8694 \\
\midrule
\multirow{2}{*}{GSM8K}
& 64  & 0.8992 & 0.9128 & 0.9151 & 0.9030 & 0.9121 & 0.8961 & 0.9037 & 0.8878 & / & / & / \\
& 128 & 0.8992  & 0.9121 & 0.9121 & 0.9121 & 0.9121 & 0.9121 & 0.9121 & 0.9030 & 0.8961 & 0.8878 & 0.8763 \\
\midrule
\multirow{2}{*}{HumanEval}
& 64  & 0.6707 & 0.6829 & 0.6829 & 0.6951 & 0.6768 & 0.6646 & 0.6341 & 0.6829 & / & / & / \\
& 128 & 0.6707 & 0.7317 & 0.7012 & 0.7012 & 0.7012 & 0.6829 & 0.6707 & 0.6707 & 0.6524 & 0.6159 & 0.6096 \\
\bottomrule
\end{tabular}
\end{table*}

\figurename~\ref{fig:cost_observation} illustrates the impact of CAEE on the system-level load imbalance by tracking the number of missed experts across different devices, a metric directly proportional to the data movement cost. The baseline configuration frequently exhibits high non-uniformity and prominent straggler devices, confirming the existence of the System-Level Imbalance (C2) problem. Crucially, the CAEE configuration actively minimizes the execution cost on the devices that represent the performance bottleneck. Across all layers, CAEE successfully reduces the transfer count on the maximum-cost device compared to the baseline, e.g., from 9 to 3 on Device 0 at Layer 30. By strategically pruning low-importance experts only when they contribute to the system's critical path latency, CAEE effectively enforces a more level cost distribution, thus directly solving the straggler problem and ensuring local savings contribute to system-wide throughput.
 
\subsection{On-device Inference}
For the on-device inference scenario, the expert importance $I(e)=\sum_{t} G(t)_e$. We then define expert-score thresholds $\theta$ to identify candidate experts $\mathcal{A}_{\text{cand}}$ for pruning. Tokens originally assigned to pruned experts are globally re-routed to higher-importance experts from $\mathcal{A}_{\text{trans}}$ via our compensation mechanism to mitigate potential accuracy degradation.

Our experiments evaluate two global concurrency levels, i.e., 64 and 128, using input length of 2k and output length of 1k data. We select $\theta$ from 0.3 and 1.2 to examine the performance-accuracy trade-off. Tables \ref{tab:on_device_performance} and \ref{tab:on_device_accuracy} summarize the latency and accuracy results, respectively.

As the expert scoring threshold increases, the token-level latency monotonically decreases, which is consistent with the cost reduction expected by our method. Regarding accuracy, we observed that in some cases, accuracy scores may slightly improve when the threshold is small. We speculate that this is because DeepSeek's group routing mechanism (designed for hardware compatibility) inherently limits routing choices; our method, however, breaks this limitation and allows for more flexible routing options that do not affect overall execution costs, potentially leading to better global utility.

Our approach maintains near lossless performance on the relatively simpler MMLU and CEval benchmarks within the valid threshold range. However, significant accuracy degradation is observed on high-complexity reasoning benchmarks like GSM8K and HumanEval when the threshold exceeds 0.7 for 64 BS and 1.0 for 128 BS. Specifically,
\begin{itemize}
\item Under a global concurrency of 64, at the $\theta = 0.6$, we achieve a $7.94\%$ improvement in end-to-end latency with no accuracy loss across all evaluated benchmarks.
\item When the global concurrency is increased to 128 and $\theta = 0.8$, the system realizes an even greater $9.59\%$ latency improvement, again with negligible accuracy degradation.
\end{itemize}

These results confirm that CAEE effectively targets and optimizes the on-device HBM pressure/memory bandwidth bottleneck, achieving substantial latency reduction while dynamically preserving model utility.

\section{Related Work}
To address the computational and memory challenges in deploying Mixture-of-Experts (MoE) models, research has primarily focused on model compression and efficient inference systems\cite{liu2024survey}. Existing related approaches can be categorized into these complementary directions: static expert pruning, dynamic routing, and expert offloading.
\subsection{Static Expert Pruning}
Expert pruning reduce model size and inference costs by eliminating parameter redundancy serving as fundamental model-level optimizations for efficient MoE deployment \cite{chen2022task,he2024demystifying,xie2024moe}. These methods mainly include structured pruning, unstructured pruning.
Structured pruning focuses on the removal or merging of entire experts. For expert removal, TSEP \cite{chen2022task} prunes non-critical experts for specific downstream tasks; NAEE\cite{lu2024not} and MoE-I²\cite{yang2024moe} eliminate redundant experts using calibration data and evolutionary search, respectively. For expert merging, DEK\cite{zhang2025diversifying} and HC-SMoE\cite{chen2024retraining} fuse experts via clustering, with the latter operating without retraining; LiteMoE\cite{zhuang2024litemoe} tailors this approach for edge scenarios by merging secondary experts to preserve core functionalities.
Unstructured pruning targets fine-grained sparsity within experts. MoE-Pruner\cite{xie2024moe} assesses weight importance by combining input activations and router weights; STUN\cite{lee2025stun} introduces a two-stage ``structured-to-unstructured'' pruning pipeline; MoE-Compression\cite{he2024demystifying} provides a unified framework integrating both paradigms.

However, these methods suffer from limitations. Structured pruning reduces expert diversity, potentially diminishing model capacity. Unstructured pruning results in irregular computation patterns that are hardware-unfriendly. Furthermore, many approaches heavily rely on calibration data or task-specific priors and incur significant offline optimization costs.

\subsection{Dynamic Routing}
Dynamic routing improve the inference efficiency of MoE models by adaptively selecting activated experts evenly\cite{aghdam2024moe,li2023adaptive,zhong2024adapmoe}.
At the routing algorithm level, research aims to design smarter gating mechanisms. DA-MoE\cite{aghdam2024moe} presents a novel method for dynamic expert allocation in MoE models, driven by token importance.\cite{li2023adaptive} adaptively allocates experts based on probability distributions, reducing computation by 38.2\%; DynMoE\cite{guo2024dynamic} proposes a top-down gating strategy for fine-grained expert assignment; XMoE\cite{yang2024xmoe} employs threshold control to dynamically activate experts. Despite their benefits, these dynamic methods typically ignore per-expert hardware cost or fail to evaluate system-wide, no-linear straggler impact.

\subsection{Expert Offloading}
Hardware-aware optimization, particularly expert offloading, leverages the sparse activation property of MoEs by caching frequently-accessed experts in HBM while evicting less active ones to DRAM or SSD, thus alleviating memory bottlenecks. To hide the associated I/O latency, prefetching and computational overlap are essential. Mixtral-Offloading\cite{eliseev2023fast} and ProMoE\cite{song2024promoe} preload experts via prediction, achieving over 90\% hit rates; Fiddler\cite{kamahori2024fiddler} and MoE-Lightning\cite{cao2025moe} leverage CPU computation to overlap with GPU execution, effectively masking load times.

Although these systems report 2$\times$–8$\times$ inference speedups\cite{kamahori2024fiddler,tang2024hobbit}, their general applicability remains questionable. They are typically evaluated in single-device or controlled settings and lack robustness in handling heterogeneous access delays and varied scenarios in production environments, thus limiting their cross-scenario optimization capability.

\section{Conclusion}
This paper introduces CAEE, a novel runtime framework designed to mitigate two critical challenges in MoE deployment: the Importance-Agnostic Cost and System-Level Imbalance. Leveraging a hardware-calibrated cost model, CAEE employs a cost-aware pruning strategy to selectively eliminate experts characterized by low token importance and high execution cost. Concurrently, a low-overhead compensation mechanism is introduced to effectively maintain model accuracy following token re-selection. Extensive experimental evaluation on the 671B DeepSeek-R1 model demonstrates the framework's efficacy across diverse scenarios. In the expert offloading scenario, CAEE reduces the TOPT by up to $17.09\%$ and the TTFT by $3.93\%$, with negligible accuracy loss. For on-device inference, it achieves a TOPT reduction of $9.59\%$, also without sacrificing utility. Furthermore, hardware observations confirm CAEE's success in directly alleviating the multi-device straggler problem. CAEE thus provides an effective, system-aware solution for scalable and efficient MoE inference.

\section*{Acknowledgements}
We acknowledge the use of AI-powered language models for proofreading, grammatical correction, and enhancing the clarity and fluency of the English language within this manuscript. We affirm that these tools were strictly used for polishing the presentation of the written work. All core intellectual content, including the foundational ideas, problem formulation, system design, experimental setup, analysis, and conclusions, remains the original and sole work of the authors.

\bibliographystyle{IEEEtranS}
\bibliography{references.bib}

\end{document}